\documentclass[]{aa}
\usepackage{graphicx}
\usepackage{txfonts}
\usepackage[longnamesfirst]{natbib}
\bibpunct{(}{)}{;}{a}{}{,}
\newcommand{\fl}{\cal F}
\newcommand{\Epk}{E_{\rm pk}}
\newcommand{\REpk}{{\cal R}E_{\rm pk}}

\defcitealias{Bor04}{B04}
\defcitealias{Fen95}{F95}

\begin{document}

\title{On the temporal variability classes found in long \\ 
gamma-ray bursts with known redshift}

\author{L.\ Borgonovo \inst{1} \and F.\ Frontera \inst{2} 
\and C.\ Guidorzi \inst{2,3,4,5}
\and E.\ Montanari \inst{2} \and L.\ Vetere \inst{6,7} 
\and P.\ Soffitta \inst{7} }

\offprints{L.\ Borgonovo, \email{luis@astro.su.se}} 

\institute{Stockholm Observatory, SE-106 91 Stockholm, Sweden
\and Dipartimento di Fisica, Universit\`a di Ferrara, 44100
Ferrara, Italy 
\and Astrophysics Research Institute, Liverpool John Moores
University, Twelve Quays House, Birkenhead CH41 1LD
\and Dipartimento di Fisica, Universit\`a di Milano-Bicocca, Italy 
\and INAF, Osservatorio Astronomico di Brera, via Bianchi 46, 23807 
Merate (LC), Italy
\and Dipartamento di Fisica, Universit\`a La Sapienza, Piazzale A.\
Moro 2, I-00185 Roma, Italy
\and INAF, IASF - Sezione di Roma, via del Fosso del Cavaliere,
I-00133 Roma, Italy
}

\titlerunning{Variability classes in GRBs}
\authorrunning{L.\ Borgonovo et al.}

\date{Received ; accepted }


\abstract{
Based on the analysis of a small sample of BATSE and Konus 
gamma-ray bursts (GRBs) with know redshift it has been reported that
the  width of the autocorrelation function (ACF) shows a
remarkable bimodal distribution in the rest-frame of the
source. However, the origin of these two well-separated ACF classes
remains unexplained.
}  
{
We study the properties of the bursts belonging to
each ACF class and look for significant differences between them.
} 
{
We complement previous ACF analysis studying the corresponding
power density spectra (PDS). With the addition of {\it Beppo-SAX \rm} 
data and taken advantage of its broad-band capability, we not only 
increase the burst sample but we extend the analysis to X-ray energies.
} 
{
The rest-frame PDS analysis at $\gamma$-ray energies shows that the
two ACF classes are not simply characterised by a different low
frequency cut-off, but they have a distinct variability as a whole in
the studied frequency range. Both classes exhibit average PDS with
power-law behaviour at high frequencies ($f' \ge 0.1$~Hz) but
significantly different slopes, with index values close to those of
Brownian ($-2$) and Kolmogorov ($-5/3$) spectra for the {\it narrow \rm}
and {\it broad \rm} classes respectively. The latter spectrum presents
an additional PDS component, a low-frequency noise {\it excess \rm}
with a sharp cut-off.  At X-ray energies we find the
power-law index unchanged for the broad class, but a significantly
steeper slope in the narrow case ($\sim -3$). We interpret this as an
indication that the broad class bursts have weaker spectral evolution
than the narrow ones, as suggested also by our analysis of the ACF
energy dependence. The low and high frequency PDS components may then arise
from two radiating regions involving different emission
mechanisms. We compare our GRB sample conditioned by
afterglow detections with a complete, flux limited BATSE sample,
finding a significant bias against narrow ACF bursts. 
} 
{} 

\maketitle




\section{Introduction} \label{intro} 

Determining the relevant timescales for any astronomical phenomenon is
essential to understand its underlying physical processes.  However,
in spite of extensive research, temporal studies on the prompt
emission phase of long gamma-ray bursts (GRBs) are not yet able to
describe and explain their basic temporal properties.  
The main challenges encountered in the temporal analysis of GRBs are
related to intrinsic characteristics of the emitted signal.  Firstly,
bursts are non-repetitive short-term events. Consequently, the total
duration of the emission in a given observational energy window is the
first timescale used to characterised them. Through out this paper we
will only consider the class of {\it long} bursts, i.e., those with
duration time $T_{90}>2$~s.  Second, burst light curves (LCs) show a
remarkable morphological diversity and they appear to have a composite
structure. While a significant fraction of bright long bursts ($\sim
15\%$) exhibits a single smooth pulse structure, in most cases they
appear to be the result of a complex, seemingly random distribution of
several pulses. Burst pulses are commonly described as having
fast-rise exponential-decay (FRED) shape, although the decay is not
strictly exponential. Therefore, the second timescale that seems
relevant for the description of a burst is a ``typical'' pulse
duration. However, analysis of the pulse parameters has shown broad
log-normal distributions not only among different bursts, but also
within a single burst \citep[see, e.g.,][]{Norr96}.
Consideration of other timescales might be relevant, e.g., for GRBs
with {\it precursors} \citep{Kos95} or when long quiescent periods
occur \citep{NP02}, although these temporal features appear only in a
small fraction of bursts.

Due to these characteristics, much of the GRB temporal
analysis has been done directly on the LCs, i.e., modelling the
pulses and studying their shape and distribution. Standard linear
analysis tools must be used with some caution, since most of the
inferences based on them would in principle require ``long'' stationary
signals, i.e., the most suitable bursts are those where the duration
is much longer than the typical pulse width.  This is the case of the
temporal analysis based on power density spectra (PDS).  Individual
PDS of GRB have very diverse shapes, and they do not seem to have
common features, although the longest bursts show spectra with a
consistent power-law behaviour.  One way to overcome these limitations
is to estimate an average PDS from a sample GRBs. This approach will
only produce physically meaningful results if each burst can be
considered a realisation of the same stochastic process, i.e., there
are no subclasses in the sample.

Under this assumption \citet{BSS98,BSS00} calculated with a large
sample of bright GRBs, an average PDS showing a clear power-law
behaviour extended over two frequency decades (approximately within
the 0.01--1~Hz frequency range), and more remarkably with an exponent
value approximately equal to that of the Kolmogorov spectrum found in
fluid turbulence \citep{Kol41}.
The significance of a thus obtained average PDS depends on two
important additional factors. First, light
curves have to be normalised to balance the weight between bursts
of different {\it brightness}. 
It is not clear at this point which norm should be used to produce the
most meaningful average. \citet{BSS98} favoured the use of the peak
flux normalisation, however they tested several other norms with
qualitatively similar results. Thus, a good normalisation should give
better convergence but the norm should not affect the final result for
a sufficiently large sample. The second problem is the shift in
frequencies produced by cosmic dilation effects. Having no redshifts
$z$ determined for their GRB sample,
\citet{BSS00} did not correct for these effects. However, they argued
that if the underlying PDS shape for every GRB is a featureless single
power-law with a constant exponent, the frequency shifts will not affect
the obtained average PDS. If this is the case, considering that the
two decades range of the power-law is much larger than the standard
deviation of the redshift distribution ($\sigma_z \sim 2$ based on the
few known redshifts), indeed the shifts should just {\it smear} the
cut-off frequencies.

The same statistical approach was used by \citet[][ hereafter
F95]{Fen95} in their study of the average autocorrelation function
(ACF) of a sample of bright GRBs. The ACF gives a measure of the
correlation between different points in the light curve that are
separated by a given time lag. Since it is the Fourier transform of
the PDS, it contains in principle the same information that can be
visualised in a different way. Therefore, the same caveats regarding
the average PDS apply to the average ACF. It was only after the
discovery of the afterglow emission \citep{Cos97} and the
determination of their redshift for a significant number of bursts
that we were able to address some of those issues. \citet[][ hereafter
B04]{Bor04} showed for a sample of 16 bright GRB with known $z$ that
when corrected for cosmic dilation effects the ACFs exhibit a clear
bimodal distribution. Using as a measure the half-width at
half-maximum, there is a highly significant gap between a narrow and a
broad width class, the separation in standard deviations being $>\!7
\sigma$. The estimated local or intrinsic values (i.e., those
calculated at the rest-frame of the source) for the average widths
were 1.6~s and 7.5~s, and the relative dispersions were 32\% and 4\%
for the narrow and broad classes, respectively. It is remarkable the low
dispersion found in the last subset, which comprised $\sim 1/3$ of the
total sample.

This article builds on the ACF analysis done in \citetalias{Bor04}.
In Sect.~\ref{data} we present our data samples and in
Sect.~\ref{methods} we briefly describe the methods used in the
subsequent temporal analysis. 
In Sect.~\ref{gLC} we strengthen previous findings based on the ACF,
expanding the previous sample of GRB with known $z$ by the inclusion
of proprietary data from the {\it BeppoSAX} mission, and we complement
the temporal analysis estimating the {\it intrinsic} PDS for each of
the subsets identified using the ACF.  Furthermore, using the
broadband capability of {\it BeppoSAX} combined instruments, in
Sect.~\ref{XLC} we are able to extend the study to the X-ray energies.
In Sect.~\ref{class}, we investigate whether the typical values of
several physical parameters commonly used to characterised GRBs differ
significantly between the two temporal classes. We look into the
problem of the energy dependence of the ACF and discuss possible
biases in our sample of bursts with known $z$ in
Sect.~\ref{tau_E}. In Sect.~\ref{discussion} we discuss our
main results.


\begin{table*}[t]
\caption{Sample of 22 GRBs with known redshift. The columns give the
name of the GRB, the source instrument for the $\gamma$-ray data, the
X-ray instrument when available, the measured redshift~$z$, the
corresponding reference, the ACF half-width at half-maximum in the
$\gamma$-ray energy band $w_{\gamma}$, the half-width corrected for
time dilation $w'_{\gamma}$, the corresponding half-widths $w_{\mathrm
X}$ and $w'_{\mathrm X}$ from the X-ray light curves when available,
the width ratio between the two energy bands $w_{\mathrm
X}/w_{\gamma}$, the index $\xi$ assuming a width energy dependence 
$w_{\gamma}(E) \propto E^{-\xi}$, and the ACF width class.}
\label{tab:sample} $$
\begin{array}{llclcccccccc}
\hline
\hline
\noalign{\smallskip}
\mathrm{GRB} & 
\mathrm{Instrument} \; (\gamma) &
\mathrm{Inst.} \; (\mathrm{X}) &
 z &
\mathrm{Ref.}^{b} & 
w_{\gamma} & 
w'_{\gamma} &
w_{\mathrm X}  & 
w'_{\mathrm X} &
w_{\mathrm X}/w_{\gamma} &
\xi_{\mathrm 2ch}^{c} &
\mathrm{Class}^{d}  \\
   &  \textrm{(55--320 keV)}^{a} &  \textrm{(2--28 keV)} & & & (s) & (s) & (s) & (s) & &  & \\
\noalign{\smallskip}
\hline 
\noalign{\smallskip}
970228 & \mathrm{ GRBM} 	&  \mathrm{WFC}    & 0.695     &   (1)	& 1.3\pm0.1	&  0.77\pm0.06  & 3.5\pm0.2  &  2.1\pm0.1  &  2.7  & 		 & \mathrm{n} \\
970508 & \mathrm{ BATSE/GRBM }  &  \mathrm{WFC}	   & 0.835     &   (2)	& 2.7\pm0.1   	&  1.47\pm0.05  & 10.3\pm0.4 &  5.6\pm0.2  &  3.8  & 0.06\pm0.05 & \mathrm{n} \\
970828 & \mathrm{ BATSE}      	&		   & 0.9578    &   (3)	& 15.33\pm0.06  &  7.83\pm0.03  &	     &	   	   &       & 0.09\pm0.03 & \mathrm{b} \\
971214 & \mathrm{ BATSE/GRBM/Konus}&  \mathrm{WFC} & 3.418     &   (4)  & 8.02\pm0.08   &  1.81\pm0.02  & 11.8\pm0.8 &  2.7\pm0.2  &  1.5  & 0.10\pm0.04 & \mathrm{n} \\
980326 & \mathrm{ GRBM }        &  \mathrm{WFC}	   & 1.2       &   (5)	& 1.34\pm0.1	&  0.61\pm0.04  & 2.35\pm0.4 & 1.05\pm0.2  &  1.7  & 		 & \mathrm{n} \\
980329 & \mathrm{ BATSE/GRBM/Konus}& \mathrm{WFC}  & 3\pm1     &   (6)	& 5.96\pm0.02   &  1.5\pm0.5    &  8.8\pm0.4 & 2.2\pm0.7   &  1.5  & 0.05\pm0.01 & \mathrm{n} \\
980425 & \mathrm{ BATSE/GRBM}  	& \mathrm{WFC}	   & 0.0085    &   (7)	& 7.62\pm0.08   &  7.56\pm0.08  & 14.8\pm0.9 & 14.7\pm0.9  &  1.9  & 0.24\pm0.06 & \mathrm{b} \\
980703 & \mathrm{ BATSE}      	&		   & 0.966     &   (8)	& 14.15\pm0.1   &  7.19\pm0.05  &	     &	    	   &       & 0.11\pm0.01 & \mathrm{b} \\
990123 & \mathrm{ BATSE/GRBM/Konus}& \mathrm{WFC}  & 1.600     &   (9)	& 19.81\pm0.03  &  7.62\pm0.01  & 35.6\pm0.2 & 13.70\pm0.08 &  1.8 & 0.30\pm0.05 & \mathrm{b} \\
990506 & \mathrm{ BATSE/GRBM/Konus}&		   & 1.3066    &  (10)	& 3.83\pm0.02   &  1.66\pm0.01  &	     &	   	   &       & 0.20\pm0.05 & \mathrm{n} \\
990510 & \mathrm{ BATSE/GRBM/Konus}&		   & 1.619     &  (11)	& 2.54\pm0.03   &  0.97\pm0.01  &	     &	   	   &       & 0.20\pm0.01 & \mathrm{n} \\
990705 & \mathrm{ GRBM }   	& \mathrm{WFC}	   & 0.86      &  (12)  & 14.3\pm0.2	&  7.7\pm0.1    & 22.3\pm0.4 & 12.0\pm0.2  &  1.6  & 		  & \mathrm{b} \\
990712 & \mathrm{ GRBM } 	& \mathrm{WFC}	   & 0.433     &  (13)	& 4.1\pm0.2	&  2.85\pm0.1   &  4.8\pm0.2 &  3.35\pm0.1 &  1.2  & 		  & \mathrm{n} \\
991208 & \mathrm{ Konus }     	&		   & 0.7055    &  (14)	& 3.67\pm0.04   &  2.15\pm0.02  &	     &	   	   &       & 		  & \mathrm{n} \\
991216 & \mathrm{ BATSE/GRBM/Konus}&		   & 1.02      &  (15)	& 3.80\pm0.02   &  1.88\pm0.01  &	     &	   	   &       & 0.18\pm0.02 & \mathrm{n} \\
000131 & \mathrm{ BATSE}      	&		   & 4.500     &  (16)	& 5.77\pm0.08   &  1.05\pm0.01  &	     &	   	   &       & 0.21\pm0.06  & \mathrm{n} \\
000210 & \mathrm{ GRBM/Konus }  & \mathrm{WFC}	   & 0.846     &  (17)	& 2.4\pm0.2   	&  1.3\pm0.1    & 5.35\pm0.3 &  2.9\pm0.2  &  2.2  & 		  & \mathrm{n} \\
000214 & \mathrm{ GRBM} 	& \mathrm{WFC}	   & 0.47      &  (18)  & 2.5\pm0.4	&  1.7\pm0.3    &  6.8\pm0.3 &  4.65\pm0.2 &  2.7  & 		  & \mathrm{n} \\
010222 & \mathrm{ GRBM/Konus }  & \mathrm{WFC}	   & 1.477     &  (19)	& 3.68\pm0.07   &  1.48\pm0.03  & 42.2\pm0.4 & 17.0\pm0.2  & 11.5  & 		  & \mathrm{n} \\
010921 & \mathrm{ GRBM	}       &		   & 0.451     &  (20)  & 9.8\pm0.3	&  6.75\pm0.2   &	     &	   	   &       & 		  & \mathrm{b} \\
011121 & \mathrm{ GRBM/Konus }  & \mathrm{WFC}	   & 0.362     &  (21)	& 10.0\pm0.3    &  7.35\pm0.2   & 18.9\pm0.2 & 13.9\pm0.15 &  1.9  & 		  & \mathrm{b} \\
030329 & \mathrm{ Konus}      	&		   & 0.1685    &  (22)	& 2.6\pm0.1     &  2.19\pm0.08  &	     &	   	   &       & 		  & \mathrm{n} \\
\noalign{\smallskip}
\hline
\end{array}
$$
\begin{list}{}{}

\item[$^{a}$] 
The energy range of BATSE data which was taken as reference instrument. 
For Konus and GRBM data the actual energy ranges are 50--200 keV and 40--700 keV 
respectively.  

\item[$^{b}$] 
(1) \citet{bloom01};
(2) \citet{mrm97};
(3) \citet{djo01};
(4) \citet{kul98};
(5) \citet{bloom99};
(6) \citet{lcr99};
(7) \citet{tin98};
(8) \citet{djo98};
(9) \citet{kul99};
(10) \citet{bloom03};
(11) \citet{bea99};
(12) \citet{Ama00};
(13) \citet{vea01};
(14) \citet{dod99};
(15) \citet{vea99};
(16) \citet{And00};
(17) \citet{pir02};
(18) \citet{Ant00};
(19) \citet{jha01};
(20) \citet{Pri02};
(21) \citet{gar03};
(22) \citet{grei03}.

\item[$^{c}$] Index $\xi$ has been estimated for an energy window width of two BATSE channels.

\item[$^{d}$] (n) and (b) indicate {\it narrow} and {\it broad} width ACF 
class, respectively.

\end{list}
\end{table*}





\section{Data} \label{data}

This work is mainly based on the analysis of light curves from GRBs
with known redshift. Given the scarce number of cases available for
study, we combined data (in the $gamma$ energy band) from three
instruments to improve our statistics. Increasing a sample in this way
presents an obvious trade-off, since we use count time series and the
difference between instrument responses introduce an additional
dispersion that may counteract the benefits. For this reason, we
initially analysed the data of the bursts that were observed by more
than one instrument, evaluating whether the differences were
acceptable for our purposes.

The comparisons were made taking BATSE as the reference instrument. Its
data comprise half of our GRB sample (in the gamma energy band),
showing the best signal-to-noise ratio (S/N) with relatively low
directional dependence thanks to the large collecting area of its
eight Large Area Detectors (LADs) placed on each corner of the {\it
Compton Gamma-Ray Observatory} \citep[CGRO;][]{Fish89}, giving full
sky coverage. It flew during the period 1991--2000 collecting the
largest GRB catalog up to date. The {\it CGRO} Science Support Center
(GROSSC) provides the so-called concatenated 64 ms burst data, which
is a concatenation of the three standard BATSE data types DISCLA,
PREB, and DISCSC. All three data types have four energy channels
(approximately 25--55, 55--110, 110--320, and $>320$ keV). The DISCLA
data is a continuous stream of 1.024 s and the PREB data covers the
2.048 s prior to the trigger time at 64 ms resolution, both types
obtained from the 8 LADs. They have been scaled to overlap the DISCSC
64 ms burst data, that was gathered by the triggered LADs (usually the
four closer to the line of sight). This combined data format was used
when available, since the concatenated pre-burst data allows a better
estimation of the background. In the case of GRB 970828 the DISCSC
data are incomplete, and we used instead the 16-channel MER data type,
binned up into 4 DISCSC-like energy channels.  All BATSE bursts with
known $z$ were considered for study, excluding two cases were
the data are incomplete or were no recorded (i.e., GRB~980326 and
GRB~980613), resulting in a total of 11 cases.

We also include the set of bursts selected in \citetalias{Bor04} that
were observed by Konus, which is a GRB detector on board the {\it
Wind} mission \citep{Apt95}. Light curves of its bursts are publicly
available at 64~ms resolution in the 50--200~keV energy band. The
collecting area of this experiment is about 20 times smaller than the
one of BATSE and consequently, in most cases, the signal is too weak
for our temporal studies. However, in
\citetalias{Bor04} a comparative ACF analysis of the bursts observed by
both instruments  showed a good agreement for very {\it bright}
bursts, and the selection criteria for Konus cases were set requiring peak
count rates larger than 3000 counts s$^{-1}$ and the availability of
post-burst data, resulting in the 5 Konus GRBs included in
\citetalias{Bor04} burst sample. 

The {\it BeppoSAX} mission, that operated between the
years 1996--2002, had broad energy band capabilities thanks to the
combined operation of several instruments. The Gamma Ray Burst Monitor
\citep[GRBM; see][]{Fro97} covered the 40--700 keV energy range, 
roughly matching the range of Konus and BATSE (i.e., 55--320 keV using
$2+3$ channels). Note that since the ACF and the PDS are quadratic
functions of the number of counts, and generally there are more counts
at lower energies, the agreement for these temporal analysis functions
will depend mainly on having a similar lower-end energy limit.  We
made use of the standard high resolution 7.8125 ms LCs, but we binned
them up into a 62.5 ms time resolution to improve the S/N ratio. This
has a negligible effect on the measurement of the time scales that
concern us here and it reduces some noise artifacts (i.e., the noise
becomes more Poissonian). In addition, it approximately matches the
standard 64~ms BATSE temporal resolution for better comparison. The
LCs were dead-time corrected and background subtracted. In
\citet{Bor05} a comparative ACF analysis was presented, including 17
GRBS detected by the GRBM with known redshift from which 8 were also
observed by BATSE. It was concluded there that although the
measured dispersions of the local ACF widths were larger than in the
BATSE case ($\sim 15\%$ at half-maximum), the average sample values
for each ACF width class were equal to within
uncertainties. Therefore, at least in the context of the present
temporal variability analysis, we can consider that these data have
more intrinsic dispersion, but the introduced errors are mainly
stochastic.


\begin{figure}
\resizebox{\hsize}{!}{\includegraphics{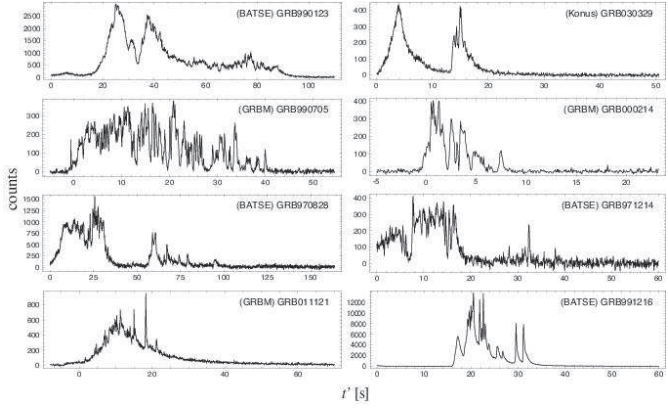}}
\caption{Examples of GRB light curves from the our sample of bursts
with known redshifts, where the time $t'$ is calculated at the source
rest-frame. The light curves on the left (right) panels have broad
(narrow) ACF widths. There are no obvious morphological differences
between the two classes, both presenting cases of simple and complex
structures.
\label{fig:LCs}}
\end{figure}


The two Wide Field Cameras (WFCs) also on board {\it BeppoSAX} covered
the 2--26 keV energy range \citep{Jag97}. During their operation time
they detected 53 GRBs in conjunction with the GRBM allowing the first
broad band studies of GRBs \citep[see, e.g.,][]{Ama02}. Furthermore, a
considerable fraction of the WFC bursts ($\sim 36\%$) were also
detected by BATSE. The LCs were extracted with a time
resolution of 62.5 ms and discriminated in three energy channels
(i.e., 2--5, 5--10, and 10--26 keV). The energy intervals were chosen
in order to have a similar amount of counts in each channel for a
typical GRB. However, except for the brightest GRBs, the signal in
each channel is too weak for the purposes of our temporal analysis and
the LCs had to be integrated into a single energy channel. Here we
will focus on the analysis of the 13 GRBs for which we
have redshift estimations. These bursts constitute a subset of our
sample of GRBs with known redshift in the $\gamma$ energy
band. Table~\ref{tab:sample} lists all these bursts indicating in its
first columns their name, the source instruments  in $\gamma$ and X-ray
bands when available, the estimated redshift $z$, and the
corresponding reference.

To further study the ACF width energy dependence we selected from the
BATSE current catalog all long bursts (i.e., duration time
$T_{90}>2$~s) with a peak flux measured on the 1.024~s timescale
$F_{\rm 1s} \ge 4$ photons ${\rm cm}^{-2} {\rm s}^{-1}$ in the
50--300~keV band that have available concatenated 64~ms data. This
resulted in a sample of 188 bright bursts for which, in most cases, the
redshift is unknown.


\section{Methods} \label{methods}

For the autocorrelation function analysis we follow the same
method presented in \citetalias{Bor04}, that was based on earlier
works of \citet{LRW93} and \citetalias{Fen95}. Here, we will summarise
the method, and we refer to \citetalias{Bor04} for further details.
Following the same notation, from a uniformly sampled count
history with $\Delta T$ time resolution and $N$ time bins, let $m_i$
be the total observed counts at bin $i$. Also let $b_i$ be the
corresponding background level and $c_i = m_i - b_i$ the net
counts. The discrete ACF as a function of the time lag $\tau = k
\Delta T$ is
\begin{equation}
A(\tau=k \Delta T) \equiv \sum_{i=0}^{N-1} \frac{c_{i} c_{i+k}-m_i \delta_{0k}}{A_0},\;\;\;\; k=0,
\ldots , N-1 \; ,
\label{acf}
\end{equation}
\noindent where $\delta$ is the Kronecker function. 
Here the periodic boundary conditions
($c_i=c_{i+N}$) are assumed. The normalisation constant $A_0$ is
defined as
\begin{equation}
A_0 \equiv \sum_{i=0}^{N-1} (c_{i}^2 - m_i) \; ,
\label{A0}
\end{equation}
\noindent such that $A(0)=1$ for $k=0$.
The term $m_i$ in Eq.~\ref{A0} subtracts the contribution of the
uncorrelated noise assuming that it follows the Poisson statistics.

For practical reasons, the actual calculation of Eq.~\ref{acf} was
done using a Fast Fourier Transform (FFT) routine. Denoting by $C_f$
the Fourier transformed of the background subtracted light curve, then
the definition of the power density spectrum (PDS) can be written as
$P_f \equiv |C_f|^2$. The noise contribution is subtracted from the
PDS assuming  Poisson statistics.
Since the PDS and the ACF are Fourier pairs (Wiener-Khinchin theorem),
the latter is obtained by inverse transforming the first. Zero padding
of the time series was used to avoid the artifacts produced by the
periodic boundary condition. 
The normalisation used for the ACF, that gives to its central maximum
unity value, is equivalent (aside from a noise correction term) to the
scaling of the LC by the square root of total power $\sqrt{P_{\rm
tot}}$, where $P_{\rm tot} \equiv \sum_{i} c_i^2$. This normalisation
is a natural choice for the ACF analysis and it makes the ACF of each
burst independent (to first order) of its brightness.  Note however
that for our PDS analysis we found it more suitable to scale the LCs
by their respective net count fluences ${\fl} \equiv \sum_{i} c_i$ (or
equivalently dividing the PDS of the original LCs by ${\fl}^2$).
Since the $a_0$ zero order Fourier coefficient is equal to the fluence
${\fl}$, our normalisation makes the PDS converge to unity towards the
low frequencies, therefore in Fourier space it can be interpreted also
as normalising by the amount of {\it power} at the lowest frequency.
In Sect.~\ref{gpds} we will further discuss this choice over other
possible normalisations.

The background estimations were done by fitting with up to a second order
polynomial the pre- and post-burst data, that were judged by visual
inspection to be inactive. This is critical for weak bursts and
particularly aggravated in the Konus case since the publicly available
LCs have almost no pre-burst data.  This is the main reason why dim bursts
were excluded in our sample selection in Sect.~\ref{data}. 

In \citetalias{Bor04} it was found empirically, considering the
analysis of the average ACFs of the narrow and broad width sets and
their dispersion, that at half-maximum the separation between the two
sets is most significant. Therefore, the ACF half-width at
half-maximum $w$ given by $A(\tau\!=\!w) \equiv 0.5$ (here we deviate
from the 
\citetalias{Bor04} notation for simplicity) was chosen as the measure 
that best characterises each class.  Since it has been shown
that the average ACF decreases approximately following a stretch
exponential \citepalias{Fen95}, the ACF width $w$ was calculated fitting the
logarithm of the ACF in the range $0.4 \le A(\tau)\le 0.6$ with a
second degree polynomial.

The uncertainties due to stochastic fluctuation were estimated using a
Monte Carlo method. For any given LC a large number of realisations is
generated assuming that the fluctuations on the gross number of counts
in each bin $m_i$ follows a Poisson distribution. Then, the LCs are
background subtracted and the ACF widths $w$ are calculated following
the same method as for the original data. However, when subtracting
the Poisson noise contribution in the ACF (Eq.~\ref{A0}) a factor of 2
must be introduced in the corresponding noise term to take into
account that the procedure to generate the synthetic LCs doubles the
noise variance, since they are in a sense second order realisations.
In general, if one considers a Poisson process with expected value
$\mu$ and iteratively assumes its counts as expected counts of another
Poisson process, after $n$ iterations the expected value is still
$\mu$, but the variance is $n \mu$ (with $n=2$ in this case).
Finally, the standard deviation of the obtained widths is used as
estimator of the uncertainty.  Nevertheless, the main source of
uncertainty in dim bursts is the background estimation, that
introduces a {\it systematic} error in the determination of the ACF
width. Therefore, the reported uncertainties should be regarded as
lower limits.  Another likely source of systematic errors is the non
uniform response of the detectors, since we do not deconvolve the
count data.


\begin{figure}
\resizebox{\hsize}{!}{\includegraphics{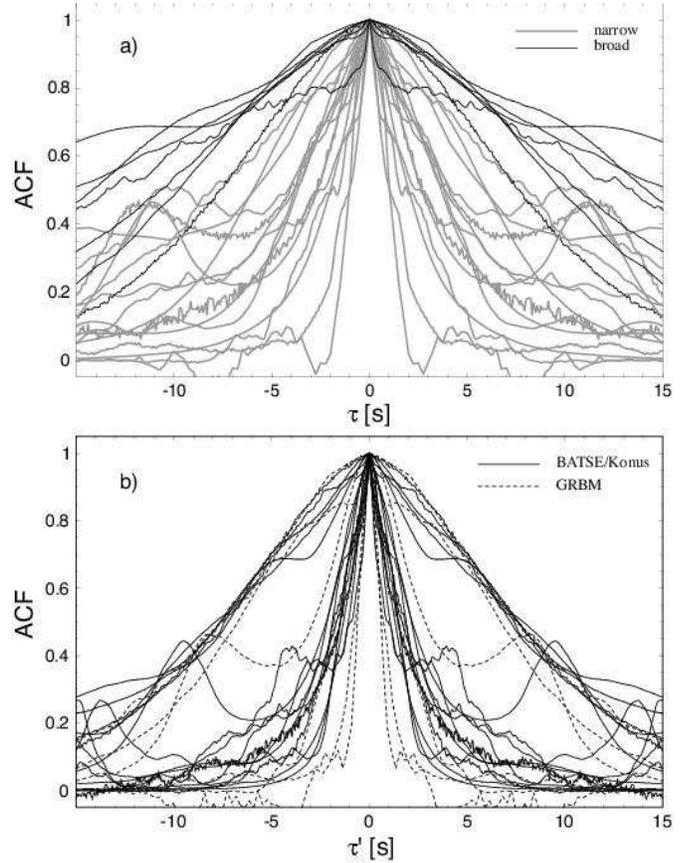}}
\caption{{\bf a)} Autocorrelation functions (ACFs) of 22 GRBs with known
redshifts calculated in the observer's frame. Although easily
identified in the rest-frame, the narrow ({\it gray lines}) and the
broad ({\it solid lines}) classes overlap giving the impression
of a unimodal distribution.
{\bf b)} Local ACFs where the cosmic time dilation effect has been
corrected, being $\tau'=\tau/(1+z)$. 
The newly added GRBM data ({\it dashed lines}) reinforce the bimodal
pattern previously found by \citetalias{Bor04} that used only the
BATSE and Konus data (here both shown with {\it solid lines}).
\label{acf_g}}
\end{figure}




\section{Analysis of the $\gamma$-ray LCs}  \label{gLC}

Some LC examples from our sample of GRBs with known redshift are
shown in Fig.~\ref{fig:LCs}. No morphological differences are evident
between the {\it narrow} and {\it broad} classes by simple
inspection. However, standard linear analysis tools reveal
clear variability differences between the two, as we report in this
section.

\subsection{ACFs} \label{gacf}

With the inclusion of GRBM bursts we were able to expand the sample
presented in \citetalias{Bor04}. The added ACFs (shown in
Fig.~\ref{acf_g} in {\it dashed lines}) also follows a bimodal
distribution when corrected for time dilation effects, reinforcing the
previously found pattern. In Fig.~\ref{acf_g}a the observed ACFs
$A(\tau)$ are shown first for comparison, while Fig.~\ref{acf_g}b
shows the rest-frame or {\it local} ACFs $A(\tau')$, where
$\tau'=\tau/(1+z)$ is the time lag corrected for cosmic
dilation. 
We obtained mean values for the rest-frame ACF half-width at
half-maximum $\bar{w'}^{\rm (n)}_{\gamma}=(1.56\pm0.15)$~s
and $\bar{w'}^{\rm (b)}_{\gamma}=(7.42\pm0.14)$~s, and also sample
standard deviations $\sigma^{\rm (n)}_{\gamma}=0.6$~s and $\sigma^{\rm
(b)}_{\gamma}=0.5$~s for the narrow and broad subsets
respectively. The mean values are equal to those reported in
\citetalias{Bor04} within uncertainties. The {\it gap}  between the two
subsets (defined as the mean difference) represents a $4.9 \sigma$
separation, smaller than the $7 \sigma$ separation reported in
\citetalias{Bor04}  mainly due to the increased dispersion in the broad 
class. However, thanks to a larger sample, we have slightly increased
the {\it significance} of this separation. The probability $p$ of a
random occurrence of such gap was estimated using Monte Carlo
methods. Assuming that there are no characteristic timescales and an
underlying uniform probability distribution (the most favourable case)
we estimated $p < 4 \times 10^{-7}$.

Hereafter our sample of bursts with known redshifts in the $\gamma$ band
will be divided for analysis into two subsets of 15 narrow and 7 broad
ACF width bursts respectively.  In Table~\ref{tab:sample} we show the
newly added ACFs widths together with those presented in
\citetalias{Bor04} for completeness.

\subsection{PDS} \label{gpds}

\begin{figure}
\resizebox{\hsize}{!}{\includegraphics{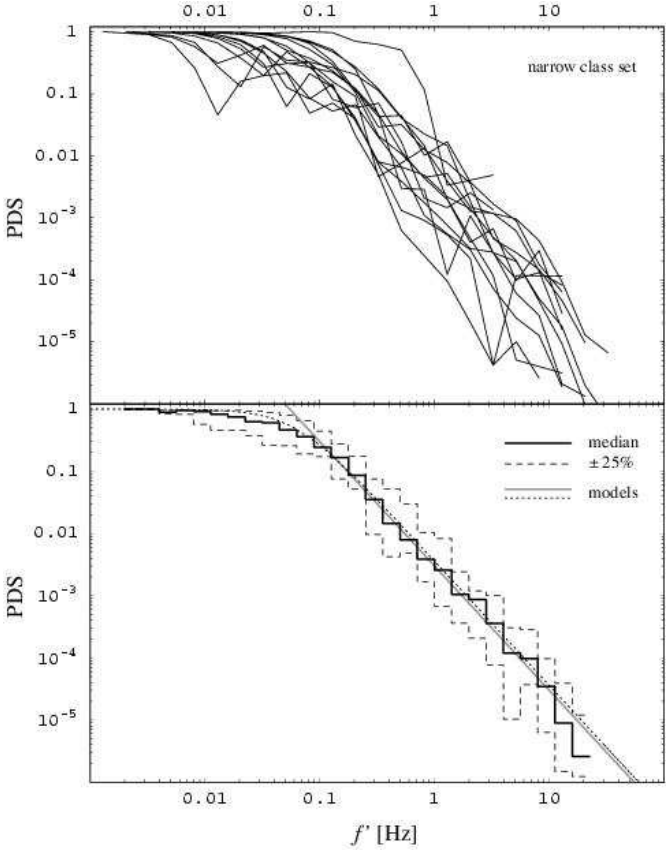}}
\caption{Power density spectra (PDS) of the subset of 15 bursts with narrow
width ACFs. The frequency scale has been corrected for cosmic time
dilation effects and the noise level has been subtracted assuming
Poisson statistics. The {\it upper panel} shows all individual PDS where
the frequency data have been equally binned in the logarithmic scale.
The {\it lower panel} shows the estimated sample median ({\it solid
line}) and the quartiles about the median ({\it dashed lines}) to
indicate the dispersion. The decay phase is well modelled by a
power-law ({\it gray line}) with index $1.97\pm0.04$. Also shown, a
Lorentzian function ({\it dotted line}) provides a fairly good fit over the
whole frequency range.
\label{PDS_narrow}}
\end{figure}

\begin{figure}
\resizebox{\hsize}{!}{\includegraphics{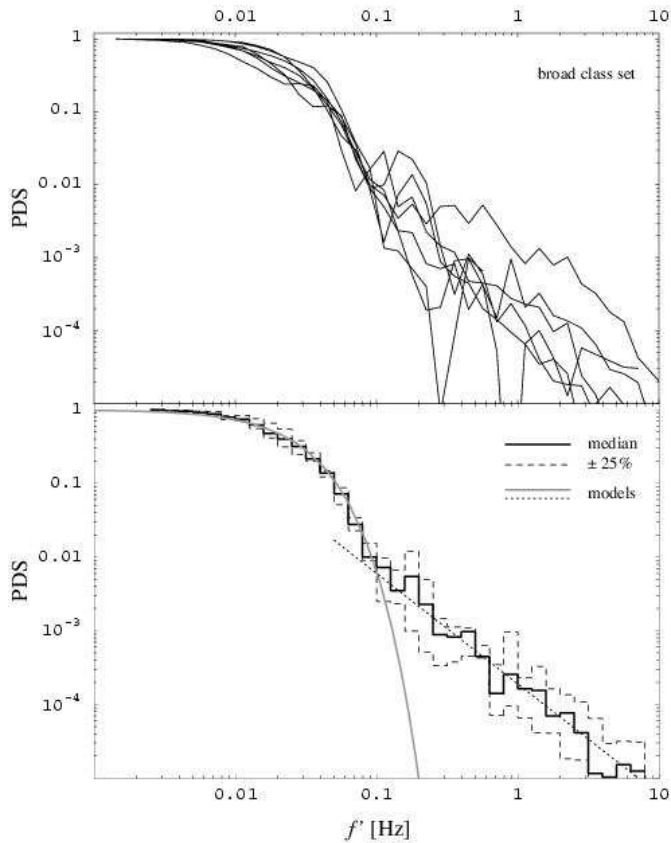}}
\caption{Power density spectra (PDS) of the subset of 7 bursts
showing broad width autocorrelation functions (ACFs). As in
Fig.~\ref{PDS_narrow} the individual PDS and the sample median are
shown for comparison. The median PDS is significantly different to
that of the narrow class and two components appear to be present. The
low frequency component has a sharp cut-off and is well modelled by a
stretched exponential ({\it gray line}). The high frequency component
is well described by a power-law with index $1.6\pm0.2$ consistent
with a 5/3 Kolmogorov spectral index.
\label{PDS_broad}}
\end{figure}

We calculate the PDS for our sample of GRBs with known redshift,
correcting the LCs for cosmic time dilation effects.  We group
bursts following the classes established in \S~\ref{gacf} based
on the ACF width. The upper panels of Figs.~\ref{PDS_narrow} and
\ref{PDS_broad} show  for the narrow and broad classes
all individual PDS overlaid for comparison. 
Given the chosen normalisation, all PDS must converge to unity at
low frequencies (see Sect.~\ref{methods}).  The frequencies have been
equally binned on the logarithmic scale. The procedure not only
smooths out the PDS stochastic variations but enable us to estimate an
average PDS for each class. This is necessary because even if all the
LCs had the same time duration, the redshift correction would make the
Fourier frequencies differ for each burst (i.e., since the
observed frequency $f$ transforms to the rest-frame as $f'=f
(1+z)$). The lower panels of Figs.~\ref{PDS_narrow} and
\ref{PDS_broad} show the corresponding central values $\tilde{P}_{f}$,
where we have chosen the use of the {\it median} over the {\it mean}
as a more robust estimator of the expected or ``underlying''
spectrum. Although qualitatively similar results are obtained in both
cases, the median shows smaller fluctuations and it is less sensitive
to the random exclusion of a few bursts from each class. The figures
show also the quartile deviation ({\it dashed lines}) as a measure of
the dispersion that was chosen to be consistent with the use of the
median (i.e., the 25\% and the 75\% quartiles around the median
value). Comparing the two median PDS it is evident that they have
remarkably different shapes. The narrow class PDS as expected shows
more power at high frequencies and it is well described by a single
power-law model $\tilde{P}_f\propto 1/f^{\alpha}$ up to a low
frequency cut-off at $\approx 0.1$~Hz. In Fig.~\ref{PDS_narrow}b we
show a fit to the median PDS in the frequency range $0.1 \le f' \le
10$~Hz for which we found a best-fit-parameter $\alpha^{\rm
(n)}_{\gamma}=1.97\pm0.04$ (i.e., $\gamma$ band - narrow class). The
obtained exponent is consistent with that of {\it Brownian or red
noise} (i.e., with $\alpha=2$), which suggests the alternative use of
a single Lorentzian model to describe the entire PDS as in {\it shot
noise} models \citep{Bel92}. However, as shown also in
Fig.~\ref{PDS_narrow}b, there seems to be a slight systematic
deviation around the cut-off region.  The broad class $\tilde{P}_{f}$
on the other hand appears to have two distinct components, i.e., a
broad low frequency component with a sharp break and a power-law
component (Fig.~\ref{PDS_broad}b).  If we assume that the two
components are independent then the single power-law should have an
independent cut-off at low frequencies which cannot be determined but
only constrained by our data.  The low frequency component is well
fitted by a stretched exponential $P_1(f)=\exp[(-f/f_1)^\eta]$. The
best-fit-parameters for this model will depend to some extent on the
cut-off frequency parameter of the power-law component and vice versa,
however this will affect mainly the estimation of the exponent
$\eta$. Taking into account these uncertainties we found a
characteristic frequency $f_1=(0.025\pm0.003)$~Hz and an index
$\eta=1.3\pm0.2$.  We considered other empirical models, concluding
that functions with power-law asymptotic behaviour (e.g., a
Lorentzian) do not decay fast enough to fit the data. For the high
frequency component we found a best-fit power-law index $\alpha^{\rm
(b)}_{\gamma}=1.6\pm0.2$ (i.e., $\gamma$ band - broad class),
consistent with the Kolmogorov spectral value~$5/3$. However, note
from Fig.~\ref{PDS_broad}a that for different bursts the latter
component appears to vary much more than the first.

\begin{figure}
\resizebox{\hsize}{!}{\includegraphics{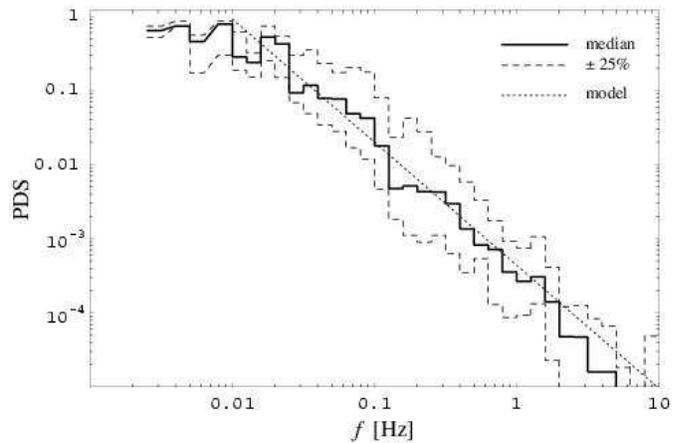}}
\caption{Median power density spectra of our whole sample of 22 GRBs with
known redshift calculated in the observer's frame. The spectrum shows
a power-law behaviour approximately within the 0.01--1 Hz frequency
range and a best fit ({\it dotted line}) gives an index
$1.73\pm0.07$, consistent with the results of  \citet{BSS98} derived 
without redshift corrections. Note that the quartile dispersion ({\it
dashed lines}) is considerably larger than those found in
Figs.~\ref{PDS_narrow} and~\ref{PDS_broad}.   
\label{mean_all}}
\end{figure}

As we discussed in Sect.~\ref{intro}, the obtained mean PDS (as well
as the median) may depend on the chosen normalisation since each burst
will be {\it weighted} in a different way. If all LCs in the sample
are realisations of the same stochastic process then the norm should
simply improve the convergence to the mean value but should give
identical results for a sufficiently large sample. Aside from the
fluence norm, we tested several ways to scale the LCs, e.g., the peak
flux used previously by \citet{BSS98,BSS00}, the \citet{Miy92}
normalisation (which expresses the PDS in fractional
root-mean-square), and the root of the total power $\sqrt{P_{\rm
tot}}$ used for the ACF. We found in all cases qualitatively the same
median PDS for each class but significantly larger spreads. 
Obviously, this does not prove that each class sample is ``uniform'' 
since it is a necessary but not sufficient condition. In this respect,
it is worth mentioning that when combined into a single sample and
without correcting for redshifts effects we obtained a median PDS
(shown in Fig.~\ref{mean_all}) that is well fitted in the 0.01--1~Hz
frequency range by a single power-law with index $1.73\pm0.07$,
consistent within uncertainties with the Kolmogorov spectrum and in
full agreement with \citet{BSS98}.
Therefore, our sample of GRBs with known $z$ is not different, at
least in this regard, to the general sample of bright BATSE GRBs used
in previous works. Note also that the overall dispersion about the
median is significantly larger than in Figs.~\ref{PDS_narrow}
and~\ref{PDS_broad} where, despite the smaller sample sizes, redshift
corrections and ACF classes have been taken into account.

\section{Analysis of the X-ray LCs}  \label{XLC}

In this section we use the WFC sample of 13 GRBs with known redshifts
(see Table~\ref{tab:sample}) to extend the previous temporal analysis
into the X-ray energy range. These bursts represent a subset of the
previously used sample in the $\gamma$-ray band (Sect.~\ref{gLC}). We
find also at these energies significant differences between the
two ACF classes although the results inevitable have larger associated
uncertainties.

\subsection{ACFs} \label{xacf}
 
\begin{figure}
\resizebox{\hsize}{!}{\includegraphics{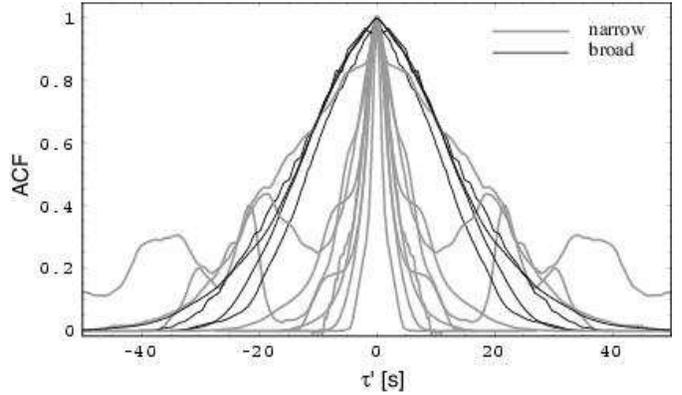}}
\caption{Autocorrelation functions (ACFs) calculated at the source
rest-frame for a sample of 13 GRBs detected by WFC in the 2--28~keV
energy band. In spite of the broadening of the ACF at lower energies,
the bimodal distribution found in the $\gamma$-ray band is clearly
seen, although with larger relative dispersion for each
class. However, one burst (GRB~010222) previously classified as
belonging to the narrow ACF width class ({\it gray lines}) appears
wider than the broad ACF width class ({\it solid lines}). This
behaviour is most likely not intrinsic but due to a systematic error
(see discussion in Sect.~\ref{tau_E}).
\label{wfc_acf_loc}}
\end{figure}

Although the ACF widths are substantially broader at these lower
energies (2--26 keV), the two ACF classes are still easily
distinguishable as shown in Fig.~\ref{wfc_acf_loc}. However, the
GRB~010222 classified as having a narrow ACF in the $\gamma$ band
({\it gray lines}) has broaded to such extent that it falls into the
broad width range, although its ACF decays more slowly than any
of the broad width cases. As shown in Table~\ref{tab:sample} where we
list all the observed and the local ACF widths, this is the only {\it
outlier} in the sample. In column 10 we list as a measure of the
broadening the ACF width ratio for the $\gamma$ and X-ray bands.
While the average width ratio is $\langle
w_{\mathrm X}/w_{\gamma} \rangle \simeq 2$, the broadening in the
GRB~010222 case ($ w_{\mathrm X}/w_{\gamma} \simeq 11.5$) is much
larger than any other burst in the sample and it is most likely caused
by a systematic error, as we will discuss later in Sect.~\ref{tau_E}.
For this reason this case will be kept separate from the sample for
the rest of our WFC data temporal analysis. Consequently, the obtained
average local width ACF are $\bar{ w'}^{\rm (n)}_{\rm
X}=(3.1\pm0.5)$~s and $\bar{ w'}^{\rm (b)}_{\rm X}=(13.6\pm0.6)$~s,
with sample standard deviations (relative dispersions) $\sigma^{\rm
(n)}_{\rm X}=1.9$~s (60\%) and $\sigma^{\rm (b)}_{\rm X}=1.6$~s (12\%)
for the narrow and broad subsets respectively.

\subsection{PDS} \label{xpds}

\begin{figure}
\resizebox{\hsize}{!}{\includegraphics{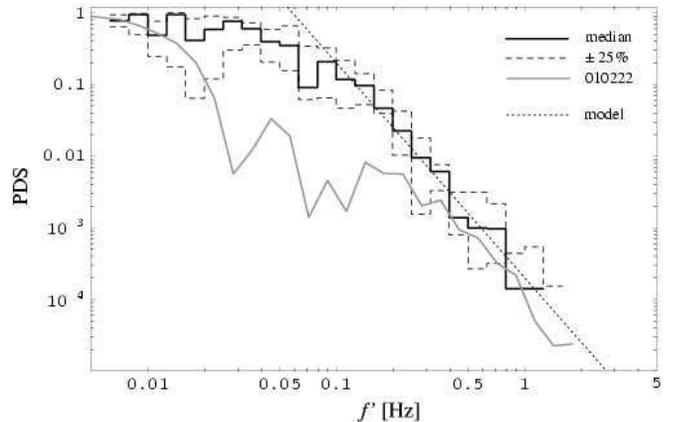}}
\caption{Median power density spectra (PDS) from the subset of WFC
bursts with known redshifts belonging to the narrow width ACFs class,
excluding the {\it outlier} case GRB~010222 that is shown along side ({\it
gray line}).  The decay at high frequencies ($f' > 0.1$~Hz)
approximately follows a power-law, but with a significantly steeper
slope than the previously found in the $\gamma$-ray energy band
($3.0\pm0.2$). GRB~010222 shows similar asymptotic behaviour at high
frequencies.
\label{wfc_PDS_n}}
\end{figure}

Once again using our sample of WFC bursts with known $z$, we
estimate the {\it local} median PDS for each ACF width class following
the same methods used in Sect.~\ref{gpds}.  We consider separately  the
special case GRB~010222 due to the unique broadening of its
ACF. Figure~\ref{wfc_PDS_n} shows our results for the narrow class. As
expected from the ACF analysis the break appears now at approximately
half the corresponding frequency  at $\gamma$ energies
(Fig.~\ref{PDS_narrow}), but most noticeably the behaviour towards
the high frequencies now shows a much faster decay. A best-fit for
frequencies $f' \gtrsim 0.08$~Hz using a $P_f\propto 1/f^{\alpha}$ model
gives an exponent $\alpha^{\rm (n)}_{X}=3.0\pm0.2$, a significantly
steeper power-law than in the $\gamma$ band. 
The PDS of GRB~010222 ({\it gray line}) is also shown in
Fig.~\ref{wfc_PDS_n} for comparison. Although with much less power at
low frequencies than the other bursts in this class, at high
frequencies ($f' \ge 0.2$~Hz) it follows a very similar asymptotic
behaviour.

\begin{figure}
\resizebox{\hsize}{!}{\includegraphics{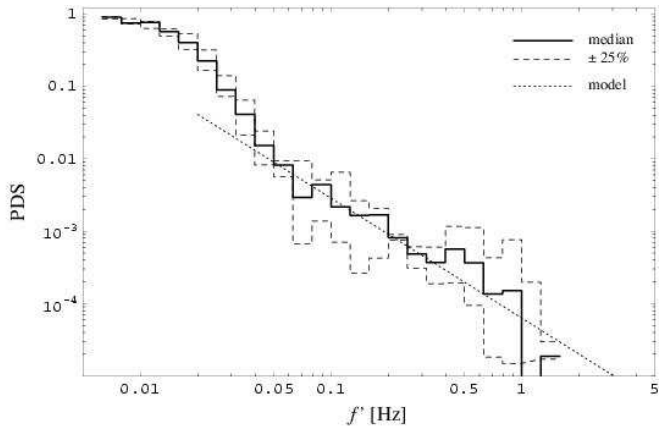}}
\caption{Median power density spectra (PDS) from the subset of 4 WFC
bursts with known redshifts belonging to the broad width ACFs
class. Despite the small sample size, the two broad PDS components
identified at $\gamma$-ray energies are recognisable. The index of the
power-law component ({\it dotted line}) is  consistent with the
Kolmogorov value as previously found in Fig.~\ref{PDS_broad}
given a best-fit-parameter $1.7\pm0.2$. 
\label{wfc_PDS_broad}}
\end{figure}

Our estimation of the median PDS $\tilde{P}_{f}$ for the broad class
is shown in Fig.~\ref{wfc_PDS_broad}. In spite of the large
uncertainties associated with such a small sample,  the two
components found in Fig.~\ref{PDS_broad} are recognisable. The low
frequency component is not as prominent and broad as before, and
consequently the data are not enough to properly constrain the fit
parameters of a stretched exponential as it was previously
done. However, we estimated that the high frequency component follows
a power-law with index $\alpha^{\rm (b)}_{X}=1.7\pm0.2$ consistent
within uncertainties with the Kolmogorov index found in
Sect.~\ref{gpds} for the $\gamma$-ray LCs. When compared with the
broad class $\tilde{P}_{f}$ in Fig.~\ref{wfc_PDS_broad}, the
GRB~010222 spectrum matches neither the $\tilde{P}_{f}$ general shape
nor any of its components. Power-law fits using the indices and the
frequency ranges shown in Figs.~\ref{wfc_PDS_n} and
\ref{wfc_PDS_broad} for the respective high-frequency components resulted in 
squared residual totals per degrees of freedom $\chi^2/\nu = 16.2/11$
and $\chi^2/\nu = 179./17$ for the narrow and broad PDS
respectively. Based on these fits we conclude that the first model is
acceptable while the second model can be rejected at a very high
confidence level. Uncertainties were estimated using synthetic shot
noise and calculating standard deviations on a sample of PDS following
the same procedures and frequency binnings applied to the burst data.

\section{Comparing ACF classes} \label{class}

In order to understand the origin of the two classes found based on
our temporal analysis we looked for any additional GRB characteristic
or physical parameter that might differ between them. 
As point out in Sect.~\ref{gLC}, the visual inspection of the LCs
reveals no trivial morphological differences between the classes (see
Fig.~\ref{fig:LCs}). Both classes present cases with simple and
complex structure, showing from a few smooth pulses to many heavily
overlapped sharp pulses.
To cover a broad range of physical parameters, we made use of the
database of GRB redshifts and other burst parameters compiled in
\citet{FB05} and references therein,
that contains much additional information about the 22 bursts in
Table~\ref{tab:sample}, although not all parameter estimations are
available for every burst. In particular, we considered the following
parameters (primed quantities are calculated at the source
rest-frame): the peak energy $\Epk'$, the isotropic equivalent
gamma-ray energy $E_{\rm iso}$, the duration time $T'_{90}$, the time
of the observed break in the afterglow light curves $T'_{\rm b}$, and
the redshift $z$.  In all cases we compared the distributions within
the narrow and broad ACF classes using the standard Kolmogorov-Smirnov
(K-S) test \citep[see, e.g.,][]{Press}, which is particularly
sensitive to median deviations, but we found no significant difference
for the cumulative distributions of any of those parameters.  No
significant $R$ correlation coefficients with $w'$ were found for
$\Epk'$ and $E_{\rm iso}$ (logarithms taken in all cases), although
marginal results were obtained for the temporal parameters $T'_{90}$
($R=0.44$ with significance $p<0.04$) and $T'_{\rm b}$ ($R=0.46$ with
significance $p<0.05$ for only 18 bursts with determined values). Note
that a correlation between $w$ and $T_{90}$ has been well established
($R=0.58$), and although in the observer-frame is partially due to the
cosmic temporal dilations, it should be found also in the rest-frame
\citep{Bor06}. However, our present sample is too small to determine
it at a high confidence level. More surprising is the finding of an
indication of some correlation between a prompt ($w'$) and an
afterglow ($T'_{\rm b}$) parameter, but this would need better
statistics to be confirmed.

Other subdivisions of the long/soft GRB class have been suggested in
the literature but we found them to be unrelated to with the one under
consideration.  Evaluation of the distributions of the observed
$T_{90}$ and $\Epk$ for the broad ACF class excludes an association
between these bursts and the ``intermediate'' duration GRB group
reported by \citet{Hor98} and \citet{Muk98}. Likewise, we found no
correspondence between the ACF bimodality and the bimodality of the
afterglow optical luminosity distribution reported by \citet{Nar06}
and \citet{LZ06}.


\section{ACF energy dependence} \label{tau_E}

In our analysis we have only corrected the LCs for cosmic time
dilation. However, the data are collected over a finite energy band
and therefore additional corrections might be needed to account for
the shift in energy, since in principle we should compare LCs emitted
in the same energy band.
As  LC pulses are sharper at higher energies
\citep{Norr96}, correspondingly the ACF is narrower.  For a sample of
45 bright bursts, \citetalias{Fen95} found that the width of the mean
ACF depends on the energy $E$ as $w(E) \propto E^{-0.43}$. Since for
large redshifts the instrument will see photons emitted at higher mean
energies, the anti-correlation of the ACF width with the energy should
partially counteract the time dilation effect. 
In \citetalias{Bor04} possible additional redshift corrections to the
accounted cosmic dilation were studied using the transformation
\begin{equation}
w'=w/(1+z)^{1+a}
\label{w'}
\end{equation}
\noindent  for the observed widths, where $a$ represents a
smaller than unity correction. The relative dispersion of each width
class was calculated as a function of the parameter $a$ looking for
minimum values
\citepalias[see Fig.~6 in ][]{Bor04} since, under the assumption that the
$w'$ intrinsic distribution does not depend on $z$ (i.e., evolution
effects are neglected), redshift dependencies should {\it increase}
the observed dispersion.  We obtained minimal dispersions at
$a^{(n)}_{\rm min}=-0.35\pm0.2$ and $a^{(b)}_{\rm min}=0.04\pm0.04$
for the narrow and broad classes respectively, equal within
uncertainties to previous results. The uncertainties were estimated
numerically using the {\it bootstrap} method \citep[see,
e.g.,][]{Press}.  If the dispersion is
calculated over the whole sample (i.e., without separating the two
width classes) no well-defined minimum is found, with an approximately
constant value within the free parameter range $|a|<1$, which further
supports the separation into two classes.  However, while the narrow
class minimum agrees within errors with the expected effect from the
energy shift, the broad class dispersion does not improve much with
any additional redshift correction.  Actually, in both cases a perfect
cancellation (i.e., $a_{\rm min}=0$) cannot be ruled out, although it
would appear  coincidental that all other possible redshift
dependencies would exactly compensate the energy shift. Another
possibility is that for the broad class the ACF dependence with energy
is weaker. At face value the low dispersion already suggests this and
in this case a nearly null $a$ correction would be much more likely.
In what follows we will show evidence that supports the latter
alternative.

\begin{figure}
\resizebox{\hsize}{!}{\includegraphics{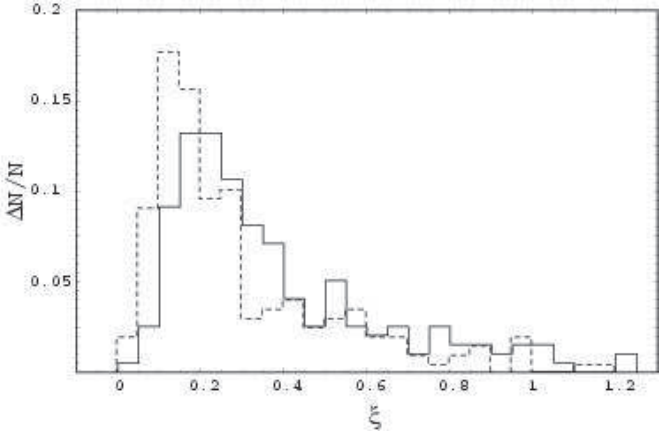}}
\caption{Distribution of the index $\xi$ in the power-law relation 
$w(E)~\propto~E^{-\xi}$ used to model the energy dependence of the ACF
width, obtained using 4-channel concatenated data ({\it solid line})
for a sample of 188 bright BATSE bursts. The distribution has an
estimated median $\tilde{\xi}_{\rm 1ch}=0.29$ and it shows an approximately
exponential decay towards larger $\xi$ values. For the purposes of
our ACF analysis that uses the two central BATSE channels, we modelled
also the $w(E)$ relation using a 2-channel energy window for each data
point. As expected, the ACF width shows a weaker energy dependence and
consequently lower typical values for the $\xi$ index. The thus
obtained distribution ({\it dashed line}) is narrower and with a
significantly lower median $\tilde{\xi}_{\rm 2ch}=0.21$.
\label{index_histo}}
\end{figure}

Using 4 energy channel data from a sample of 188 bright BATSE bursts
we analyse the ACF width dependence with energy for each individual
case using a single power-law model $w(E)~\propto~E^{-\xi}$ (the
channels are approximately equally spaced on a logarithmic scale). We
found that in most cases the relation $w(E)$ in individual bursts is
not well fitted by a power-law (less than 33\% of the bursts satisfied
for the sum of the reduced residuals $\chi^2_{\rm r}<2$), but for most
cases it can be considered a good approximation within the BATSE
energy range with typical total relative errors $<10\%$. The obtained
distribution of $\xi$ indices is shown in
Fig.~\ref{index_histo}. Since we found for our fits poorer
figures-of-merit (i.e., $\chi^2_{\rm r}$) for larger $\xi$ values, the
largest values are related to the cases that strongly deviate from
the power-law model for $w(E)$ and are therefore ill determined. 
For this same sample, a power-law fit of the energy dependence of the
width of the {\it mean} ACF gives an index $-0.40\pm0.03$ in agreement
with \citetalias{Fen95}. However, note in Fig.~\ref{index_histo} that for
such a long-tailed distribution most individual bursts typically have a
weaker energy dependence and therefore the median $\tilde{\xi}_{\rm
1ch}=0.29$ is a more representative value of individual indices.

In addition, since we are particularly interested in quantifying the
effect of the energy dependence in our ACF analysis where the
observation energy window has the width of two BATSE channels, we
studied the ACF width energy dependence $w(E)$ combining the 4-channel
data into three overlapped 2-channel width energy bands.  The
distribution of the thus obtained indices $\xi_{\rm 2ch}$ is shown
in Fig.~\ref{index_histo} ({\it dashed lines}). The median
$\tilde{\xi}_{\rm 2ch}=0.21$ is significantly smaller, as expected
since less correction will be needed the wider the energy window. If
we consider the median value of the ACF broadening $w_{\mathrm
X}/w_{\gamma}$ for the sample in Table~\ref{tab:sample} (excluding
GRB~010222) we derive a median index $\tilde{\xi} \approx 0.20$, in
good agreement with the 2-channel energy window distribution in
Fig.~\ref{index_histo}, either taking the lower edge of each energy
window \citepalias[following][]{Fen95} or their geometric means.
In Table~\ref{tab:sample} we list our estimations of
the index ${\xi}_{\rm 2ch}$ for the 11 BATSE GRBs with known $z$.
Both ACF width classes show typical values that seem consistent with
the general distribution in Fig.~\ref{index_histo}, but unfortunately
the small sample size  only allowed us to distinguish very
substantial differences.

Since we do not have redshifts for most of the BATSE GRBs, the
estimated distribution of index $\xi$ was derived mixing both ACF
width classes; nevertheless it is probably fairly representative of
the narrow class that dominates the sample. In principle, we do not
know how representative that general distribution is of the broad
class. For our sample of 11 BATSE GRBs with known $z$ the fraction of
broad cases is $f_{\rm b}=4/11$ ($f_{\rm b}=7/22$ for our whole
sample). However, since the afterglow localisations needed for the
redshift determinations always involved another instruments with
different trigger responses, that fraction is most likely not
representative of the general BATSE catalog.  An analysis of the ACF
widths of the 188 brightest long BATSE bursts sample shows that the
distribution is well described by a log-normal distribution with a
mean $10^{\langle \log w \rangle}=3.0$~s and a standard deviation
representing a factor of $3.1$, with no significant hint of an underlying
bimodality \citep[see also][]{Bor06}. Since all 11 BATSE GRB with
known redshift have $w>2.5$~s almost half of the BATSE $w$
distribution without redshifts seems truncated. Using the K-S test
that compares cumulative distributions we calculate a chance
probability $p<0.02$ that they are drawn from the same distribution.
We conclude that the sample with known $z$ is significantly biased
towards large ACF widths, mainly  affecting the representation of the
narrow class. Taking into account this truncation and the fact that for
a burst to belong to the broad class its observed ACF width must be
$w\gtrsim 7$~s independently of the redshift, we estimate that the
unbiased BATSE fraction of broad cases should lie approximately
between $0.08 \lesssim f_{\rm b} \lesssim 0.22$ to be consistent with the
observations.

Having characterised the energy dependence of the timescale $w$ with
the index $\xi$, we investigated possible correlations with other GRB
parameters, but bearing in mind that the largest $\xi$ values are most
likely meaningless.  We tested a broad range of temporal and spectral
parameters (e.g., duration time $T_{90}$, emission time, ACF width
$w$, time lag between energy channels, $\nu F_\nu$ peak energy $E_{\rm
pk}$, low and high spectral indices $\alpha$ and $\beta$) following
definitions and methods described in \citet{Bor06} for a multivariate
correlation analysis. In particular, we used a measure of the overall
spectral evolution of a burst introduced  there, where the LC is
divided in two fluence-halves and the peak energies of the
corresponding integrated spectra are estimated and compared. In that
way a ratio of peak energies is defined as $\REpk \equiv
{\Epk}^{(1)}/{\Epk}^{(2)}$, so that $\REpk>1$ implies an overall hard
to soft evolution. Although we found no significant global
correlations with the index $\xi$, there are indications of a
different behaviour at low $\xi$ values in two of the studied
parameters, i.e., $w$ and $\REpk$.

\begin{figure}
\centering
\resizebox{\hsize}{!}{\includegraphics{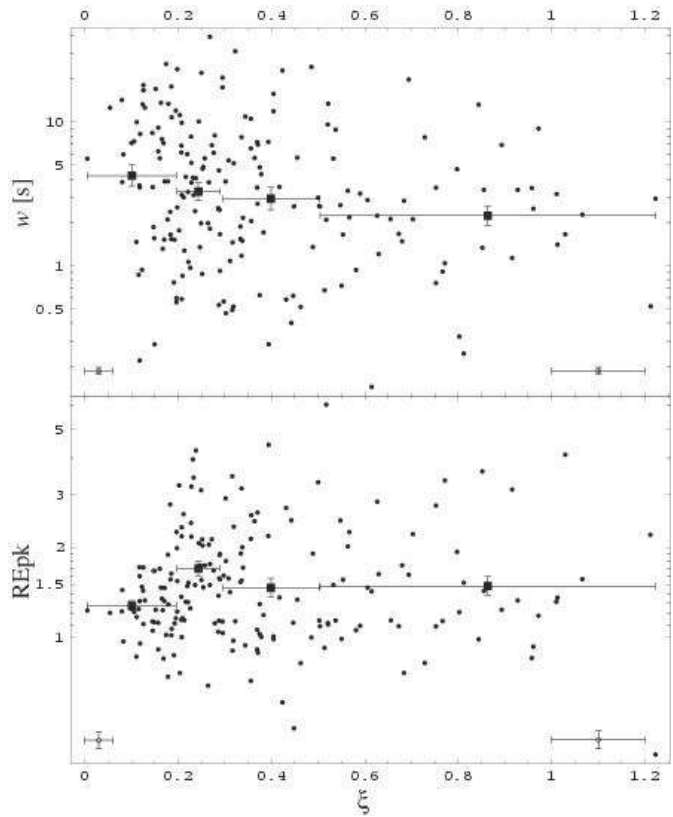}}
\caption{Correlation analysis for the index $\xi$
from the relation $w(E)~\propto~E^{-\xi}$ describing the energy
dependence of the ACF width based on a sample of 188 BATSE
bursts. Although we found no overall correlation between $\xi$ and
other commonly used burst parameters (see Sect.~\ref{tau_E}), at small
$\xi \lesssim 0.3$ values two parameters show a significantly
different behaviour comprising $\sim 50\%$ of the sample.  At the
lower corners we show average values of the estimated uncertainties
representative of each data halves. Large $\xi$ values have much
larger uncertainties and they are usually associated with poor
fits. The $\xi$ axis has been divided into four bins with equal number
of data points for averaging ({\it solid squares}).  {\it Upper
panel:} Typical values of the ACF width $w$ are higher for lower $\xi$
indices. {\it Lower panel:} The ratio of peak energies $\REpk$, that
is a measure of the burst energy spectral evolution, shows less
dispersion and a positive correlation at low $\xi$ indices.
\label{xi_corrs}}
\end{figure}

In Fig.~\ref{xi_corrs} we show scatter plots of the ACF width $w$ and
the ratio of peak energies $\REpk$ versus the index $\xi$. The $\xi$
observed range was divided into four bins with equal number of data
points, and the corresponding geometric means (more appropriate for
log-normal distributions) and uncertainties are shown. Due to the
large spread the variables seem uncorrelated based on their
correlation coefficients $R$, nevertheless the binning reveals a trend
where lower $\xi$ values correspond on average to broader ACF widths
$w$ and lower values of the $\REpk$ ratio. In addition, a weak but
significant anti-correlation exists between $w$ and $\REpk$ with
coefficient $R=-0.25$ taken their logarithms.
For $\xi\lesssim0.3$ there is noticeably less dispersion in $\log
\REpk$ (i.e., less relative dispersion in $\REpk$) and a clear
positive correlation that comprises half of the sample (with
coefficient $R=0.45$). Comparing a linear fit to those data to the no
correlation null hypothesis gives for the F-test (the standard test
for a ratio of $\chi^2$ values) a level of rejection
$p<0.001$. However, since we do not have an objective criterion to
truncate the index $\xi$ distribution, the real significance of this
found correlation is hard to evaluate; therefore we approach the
problem using Monte Carlo methods. We draw random samples for both
parameters using the bootstrap method so that they are
uncorrelated. Then sorting the data based on their $\xi$ values, we
look for the $\xi$-truncate that would give the maximum chance
correlation with $\REpk$. Assuming conservatively that the correlation
has to comprise at least 1/3 of the sample and  $|R|>0.3$ then
the trial is judged as successful. In this way we found a probability
of $p<0.008$ of finding a similar or stronger correlation by chance.
Although the energy dependence of the ACF is clearly a consequence of
the spectral variability, the spectral evolution of a burst is as
complex as its LC, shifting from hard to soft over each individual
pulse \citep{BR01}. For this reason it is only under certain limited
conditions that we are able to actually observe a correlation between
$\xi$ and $\REpk$.  In summary, from the analysis of
Fig.~\ref{xi_corrs} there are clear indications that bursts with broad
ACF on average show weaker ACF energy dependence and less spectral
evolution.

Finally, we attempted to extend the analysis of the ACF energy
dependence to X-ray energies with the use of 3-channel data from the
WFC bursts which were simultaneously observed by BATSE.  However, only
the signal of the very brightest of the WFC bursts were found suitable
for this purpose and the resultant sample was considered too small for
a systematic broad-band analysis. Furthermore, although for most of
these bursts $w(E)$ shows a smooth, approximately power-law behaviour,
we observed in a significant number of bursts a function discontinuity
or ``shift'' between the WFC and BATSE data, a systematic error most
likely due to the background estimation. The observed fraction of
cases that present this artifact ($\lesssim 20 \%$) is consistent with
the occurrence of one such case within the 13 WFC bursts with known
$z$ in Table~\ref{tab:sample}.  Therefore, we conclude that the
unusual ACF broadening of the {\it outlier} case GRB~010222 discussed
in Sect.~\ref{xacf} is most likely an artifact. The PDS analysis also
supports this since an incorrect background subtraction would mainly
affect the power at low frequencies, and we have shown in
Fig.~\ref{wfc_PDS_n} that GRB~010222 deviates from the narrow class
typical behaviour only at those frequencies.

\section{Discussion}  \label{discussion}

The reported ACF bimodality in \citetalias{Bor04} was based on the
analysis of a small sample of GRBs with available $z$. Nevertheless,
the gap between the two ACF classes was found to be highly significant
(with $p<6 \times 10^{-7}$).  Low-number statistics can often suggest
nonexistent features and should be regarded with caution.  As always,
one should consider the possibility of a selection effect not
accounted for in the probability estimation. However, while it is easy
to conceive of upper and lower biases in the observed ACF width
distribution (e.g., due to trigger criteria), it is harder to think of
a reason why this would happen for middle range $w'$ values. We
have shown in Sect~\ref{tau_E} that, when compared to a complete, flux
limited large sample of BATSE bursts, our sample with known $z$ is
likely biased against very narrow ACFs.  The addition of GRBM data
with their larger intrinsic dispersion only slightly increased the
statistical significance of the found ACF bimodality. Nevertheless, with
three sets of $\gamma$-ray data showing the same bimodal pattern
(Fig.~\ref{acf_g}) we can safely rule out an instrumental effect as
the cause of it.

It is clear from the PDS analysis presented in Sects.~\ref{gLC} and
\ref{XLC} that narrow and broad ACF bursts have very different overall
variability, with the latter showing a prominent low-frequency PDS
component with a very sharp cut-off, which explains in part the
associated low dispersion observed in this  ACF width class, despite
the typical fluctuations in individual burst PDS and the observed
dispersion in the high-frequency power-law component.
Another possible difference between the two ACF classes is hinted at by
the trends shown in Fig.~\ref{xi_corrs}, which could be interpreted as
an indication that broad ACF bursts show less spectral evolution than
the narrow ones. This view is indirectly supported by the comparison
of the respective median PDS $\tilde{P}_{f}$ at different energies
estimated in Sects.~\ref{gpds} and \ref{xpds}. The narrow class
exhibits a single power-law PDS with index
$\alpha^{(n)}_{\gamma}\approx2$ at high energies and a steeper
$\alpha^{(n)}_{X}\approx3$ low energies.  In {\it shot noise} models
where LCs are generated by random pulses, an index 2 commonly appears
when the pulses have sharp rising phase, like the typical FRED-like
pulses in many GRBs. But due to the hard-to-soft evolution
\citep{BR01} the same pulses at X-ray energies look smoother with
slower rise phases giving steeper indices.  On the other hand, the
high-frequency power-law component in the broad class PDS shows
approximately the same index in both energy bands (close to the $5/3$
Kolmogorov index). Visual inspection in the $\gamma$ and X-ray bands
of the bright broad ACF case GRB~990123 \citep[see, e.g.,][]{Cor05},
which shows well-separated pulses, indicates that the broadening of
the ACF at low energies is not as much due to a corresponding
smoothing of individual short pulses as to a decrease in their
relative importance with respect to a long smooth LC component that
seems to lie underneath them.  If this case were representative of the
whole class then that would explain in terms of LC components the
observed $\tilde{P}_{f}$ behaviour at different energies.

One remarkable feature of the found ACF bimodality is the low relative
dispersion of the broad class ($\sim 6\%$ in the $\gamma$ band) around
the $\bar{w'}^{\rm (b)}_{\gamma}=(7.42\pm0.14)$~s timescale. If
confirmed such a temporal feature would have very important
implications, both for our physical understanding of GRBs and because
of its potential use in cosmological studies. A crucial issue then is
to properly account for redshift dependencies. The reported ACF widths
were derived taking into account only the cosmic time dilation, since
other effects are of a lesser order. The latter were quantified by
means of the free parameter $a$ included in Eq.~\ref{w'} width
transformation. We found well-defined minima at $a^{(n)}_{\rm
min}=-0.35\pm0.2$ and $a^{(b)}_{\rm min}=0.04\pm0.04$ for the narrow
and broad ACF class respectively \citepalias[in agreement
with][]{Bor04}. Since \citetalias{Fen95} reported a considerable
energy dependence for the average ACF ($w(E) \propto E^{-0.43}$) then
the cosmic energy shift would be able to account within errors for the
found $a_{\rm min}$ in the narrow class. However, for the broad class
either we would need to assume that there are other dependencies at
work that coincidentally cancel out almost exactly the energy shift
effect or that simply the ACF energy dependence is weaker in this
case.  From the analysis of this alternative presented in
Sect.~\ref{tau_E} we conclude that in the context of
our ACF study the power-law energy dependence is better characterised
by a median index $\tilde{\xi}_{\rm 2ch}=0.21$, still consistent with
the narrow class within errors and a considerably smaller effect to
account for.  We evaluated by Monte Carlo simulations that the
estimated $w'$ dispersion (for a broad class sample of only 7 bursts)
is smaller but not significantly different to the one expected
assuming the ${\xi}_{\rm 2ch}$ distribution found in
Fig.~\ref{index_histo}.  Moreover, we found indications that the broad
ACF class may have lower typical $\xi$ indices than the general
sample. This would be an expected consequence of the weaker spectral
evolution already mentioned above when comparing the two classes.

Based on the analysis of WFC X-ray data, \citet{Vet06} reported that in
a fraction of bursts the existence of two LC components is
apparent. They selected 10 bursts by visual inspection and found for
their {\it slow/fast} (low/high frequency variability) LC
decomposition that the slow components were in general softer than the
corresponding fast ones, suggesting different underlying emission
processes. There are only 3 GRBs that they claim to have slow
components among our sample with known $z$, but they all show broad
ACFs in the source rest-frame (including the outlier GRB~010222),
which suggests a connection to  our own results that  will
need further analysis to be established. The discussion
about the nature of the slow/fast components in \citet{Vet06} would
apply also to the low/high frequency components found in broad ACF
bursts.  
Within the framework of the internal shock model developed by
\citet{MR00},  there would be two major radiating regions in the
relativistic outflow and the PDS components would arise from the
different expected variability of the emission, i.e., from a thermal
photospheric component and from the non-thermal flux of  optically thin
dissipative regions above the photospheric radius. In this context,
ACF class differences could be due to the conditions controlling the
relative strength of those two emission components \citep[][]{Ryde06}.

\begin{acknowledgements}  
We thank S.\ Larsson and C.-I.\ Bj\"ornsson for useful comments and
careful reading of the manuscript. We are grateful to the anonymous
referee for constructive criticism.  This research has made use of
BATSE and Konus data obtained from the High Energy Astrophysics
Science Archive Research Center (HEASARC), provided by NASA's Goddard
Space Flight Center.
\end{acknowledgements}




\end{document}